%% file: main.tex
\documentclass[conference]{IEEEtran}
\IEEEoverridecommandlockouts
\usepackage{cite}
\usepackage{kotex}\usepackage{cite}
\usepackage{amsmath,amssymb,amsfonts}
\usepackage{algorithmic}
\usepackage{graphicx}
\usepackage{textcomp}
\usepackage{xcolor}
\usepackage{hyperref}
\usepackage{graphicx}
\usepackage{booktabs}
\usepackage{bm}
\usepackage{multirow}
\input{abbreviation}
\input{helper-command}
\debugtrue

\def\BibTeX{{\rm B\kern-.05em{\sc i\kern-.025em b}\kern-.08em
    T\kern-.1667em\lower.7ex\hbox{E}\kern-.125emX}}
\begin{document}

\input{Sections/0_1_Title}
\input{Sections/0_2_Authors}
\maketitle
\input{Sections/0_Abstract}

\input{Sections/1_Introduction}
\input{Sections/2_Methods}

\input{Sections/3_Experiments}

\input{Sections/4_Results}
\input{Sections/5_Conclusion}

\input{Sections/6_Acknowledgment}

\clearpage
\bibliographystyle{IEEEtran}
\bibliography{bibliography}

\end{document}

%% file: abbreviation.tex
\newcommand{\sysname}{Maestro-EVC}

\newcommand{\spkenc}{EISE}

\newcommand{\conemo}{TCEM}
\newcommand{\emopro}{EEPT}

%% file: helper-command.tex
\newif\ifdebug
\newcounter{commentcounter}
\newcounter{unknowncounter}
\newcounter{needchartcounter}

\newcommand{\needchart}[1]{\protect\stepcounter{needchartcounter}\textcolor{blue}{Fig. XX}}

\newcounter{mycounter}

%% file: Sections/0_1_Title.tex
\title{Maestro-EVC: Controllable Emotional Voice Conversion Guided by References\\ and Explicit Prosody\\
\thanks{\textsuperscript{*}Equal contribution. Authors are listed in alphabetical order.}
\thanks{\textsuperscript{\dag}Corresponding author.}

}

%% file: Sections/0_2_Authors.tex

\makeatletter
\renewcommand{\footnoterule}{%
  \kern -3pt
  \hrule \@width \columnwidth \@height 0.4pt
  \kern 2.6pt}
\makeatother

\author{
    \IEEEauthorblockN{
        \textsuperscript{1*}Jinsung Yoon,
        \textsuperscript{1*}Wooyeol Jeong,
        \textsuperscript{2}Jio Gim,
        \textsuperscript{1,2\dag}Young-Joo Suh,
    }
    \IEEEauthorblockA{
        \textsuperscript{1}\textit{Graduate School of Artificial Intelligence}\qquad
        \textsuperscript{2}\textit{Dept. of Computer Science and Engineering}\\
        Pohang University of Science and Technology (POSTECH), Pohang, Republic of Korea\\
        \texttt{\{truestar2001, jungwy0106, jio.gim, yjsuh\}@postech.ac.kr}
    }
}

%% file: Sections/0_Abstract.tex
\begin{abstract}

Emotional voice conversion (EVC) aims to modify the emotional style of speech while preserving its linguistic content. In practical EVC, controllability, the ability to independently control speaker identity and emotional style using distinct references, is crucial. However, existing methods often struggle to fully disentangle these attributes and lack the ability to model fine-grained emotional expressions such as temporal dynamics. We propose Maestro-EVC, a controllable EVC framework that enables independent control of content, speaker identity, and emotion by effectively disentangling each attribute from separate references. We further introduce a temporal emotion representation and an explicit prosody modeling with prosody augmentation to robustly capture and transfer the temporal dynamics of the target emotion, even under prosody-mismatched conditions. Experimental results confirm that Maestro-EVC achieves high-quality, controllable, and emotionally expressive speech synthesis.

\end{abstract}


%% file: Sections/1_Introduction.tex
\section{Introduction} 
Emotional voice conversion (EVC) aims to transform a given utterance into a different emotional style while preserving the linguistic content~\cite{zhou2022emotional}. 
EVC has gained prominence due to its high potential in various applications, such as digital avatars~\cite{hussain2022training}, virtual assistants~\cite{chatterjee2021real}, and human-computer interaction~\cite{pittermann2010handling, erol2019toward}.

Practical EVC systems require two key capabilities. 
The first is controllability, which refers to the ability to control content, speaker identity, and emotional style independently.
The second is the ability to convey fine-grained emotional expressions, including temporal dynamics.
In particular, scenarios such as emotional dubbing, where generating fine-grained emotional expressions in the target voice is required, demand both controllability and emotional expressiveness.

\begin{figure}[!t]
    \centering
    \includegraphics[width=0.48\textwidth]{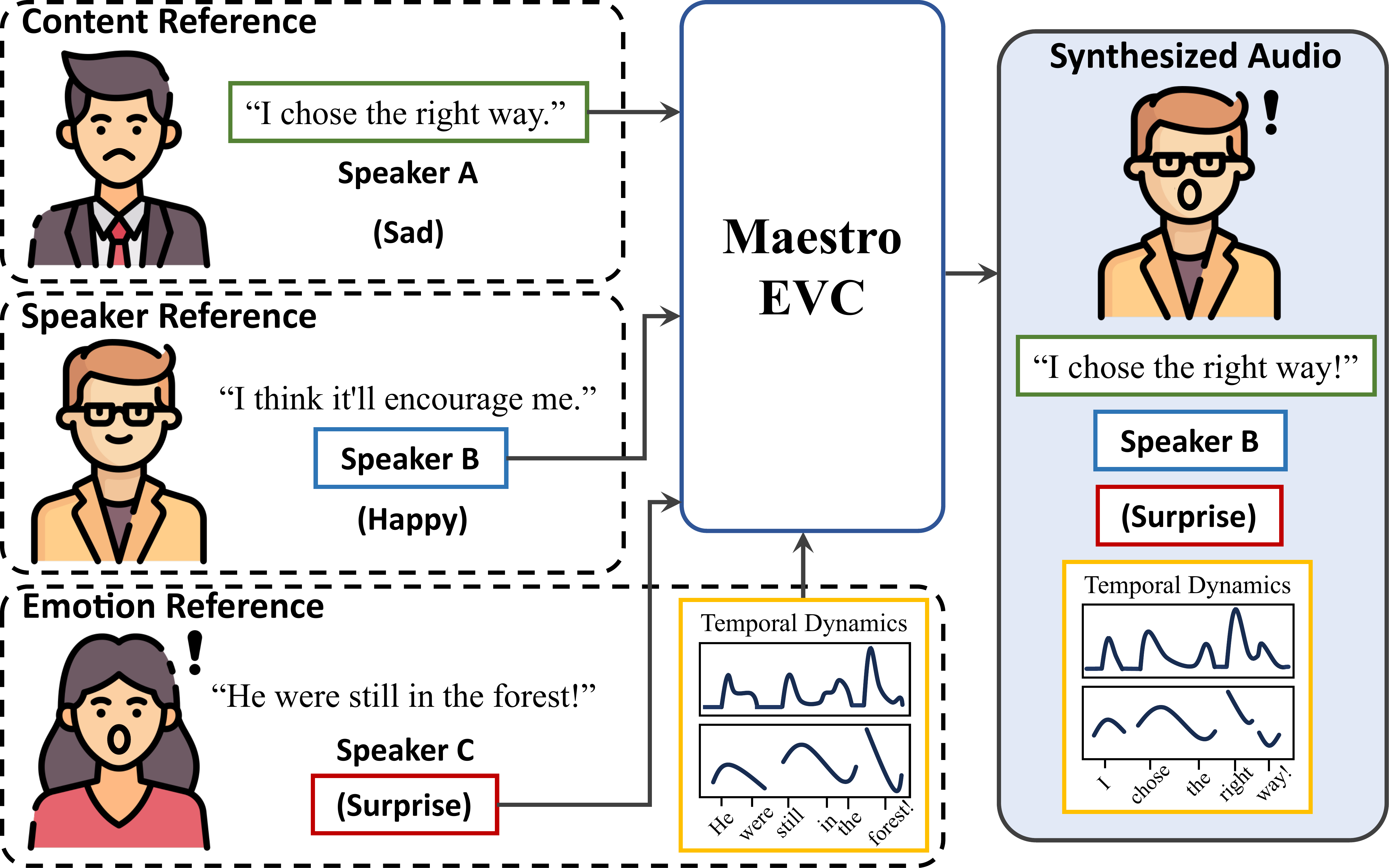}
    \caption{
        An example of speech conversion using Maestro-EVC, harmoniously integrating content, speaker identity, emotion, and temporal dynamics.
    }
    \label{fig:example}
    \vspace{-1.5em}
\end{figure}
Several approaches have been proposed to independently convert both the emotional style and speaker identity.
Some of these methods rely on predefined emotion categories (e.g., “happy,” “sad”)~\cite{qi2024pavits, zhou2020converting, zhou2021limited}, instead of using utterance-level emotion embeddings extracted from an emotion reference~\cite{zhou2022emotion, zhou2021seen, zhu2023emotional, chen2022speaker}. 
However, the use of emotion categories limits generalization to unseen emotion states and lacks the expressiveness for fine-grained emotion modeling.
Similarly, approaches that use predefined speaker IDs as input often struggle to generalize to unseen speakers.
To overcome these limitations, recent frameworks adopt fully reference-guided mechanisms that enable independent control of content, speaker, and emotion by directly conditioning on reference utterances~\cite{shah2023nonparallel, wang2025enhancing, dutta2024zero}.

\begin{figure*}[!t]
    \centering
    \includegraphics[width=1.0\textwidth]{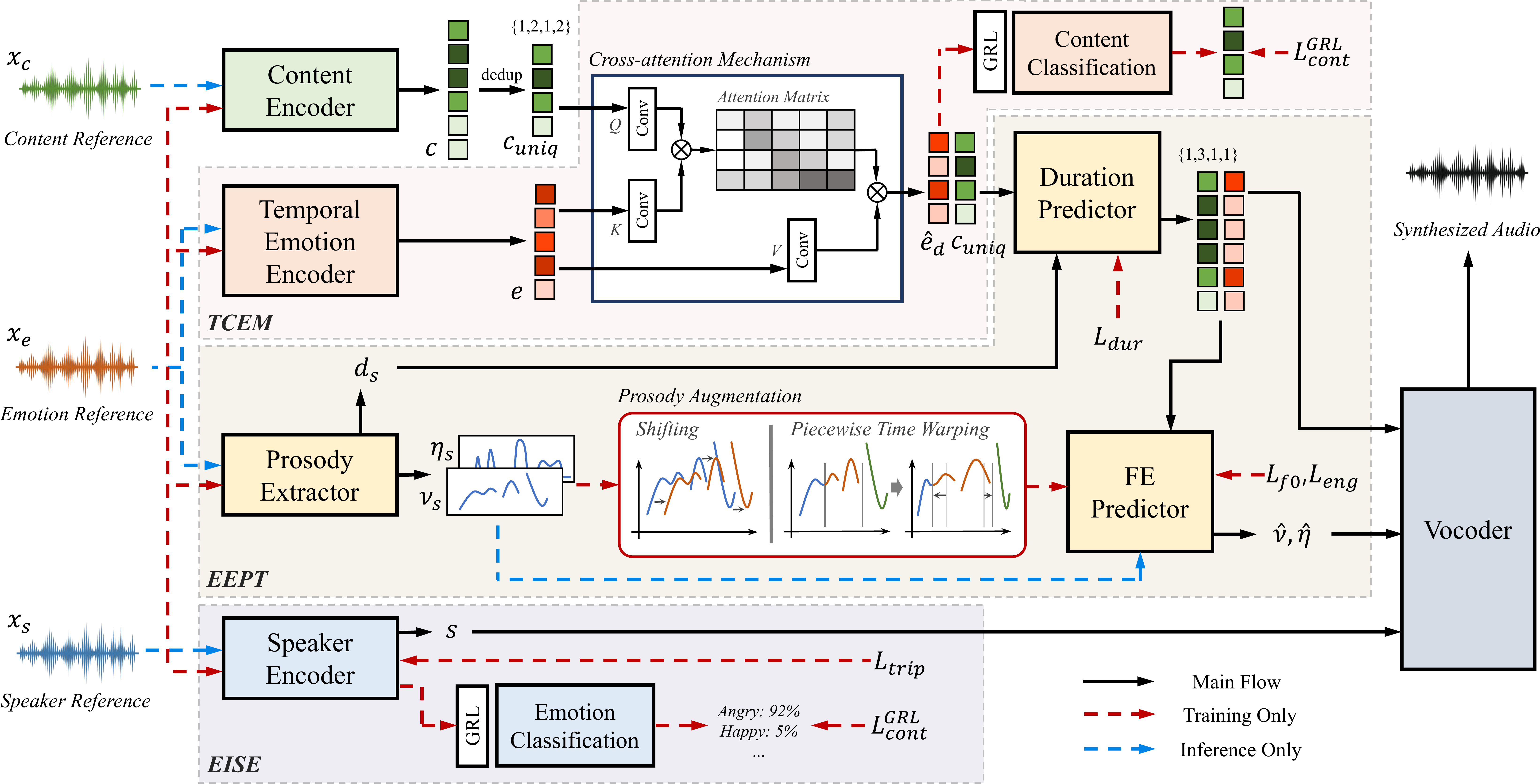} 
    \caption{
        Overview structure of the proposed {\sysname}. $\mathit{x_c}$, $\mathit{x_e}$, and $\mathit{x_s}$ denote the content, emotion, and speaker reference utterance, which are identical during training such that $\mathit{x}_c = \mathit{x}_e = \mathit{x}_s$, where the reference is a single utterance from the training dataset. This condition is illustrated by the red dashed line.
        }
    \label{fig:example}
    \vspace{-1.5em}
\end{figure*}

Among such approaches, most adopt a reconstruction-based framework~\cite{dutta2024zero, wang2025enhancing} by disentangling content, speaker, and emotion representations from a single utterance and reconstructing speech from them.
Although such approaches often produce natural speech, they struggle to fully disentangle these attributes, limiting the model’s ability to control each factor independently.
Moreover, since these methods rely on utterance-level emotion representations, they fail to capture fine-grained temporal dynamics in the emotional expression.


To effectively transfer the fine-grained temporal dynamics of the emotion reference, it is essential to extract temporal emotion representations.
For this purpose, several EVC approaches have proposed modeling prosody, such as pitch (F0), energy, and rhythm, which serve as effective carriers of temporal emotional characteristics~\cite{qi2024pavits, dutta2024zero, lu2021multi}.
Nevertheless, these studies model prosody implicitly, predicting prosodic patterns from latent representations rather than directly conditioning on prosody extracted from audio, which limits their ability to transfer fine-grained temporal dynamics.
Thus, an explicit prosody modeling strategy that conditions on actual prosody extracted from an emotion reference is required to more accurately transfer fine-grained temporal dynamics.
However, one key challenge in applying this strategy is the prosody mismatch between the emotion and content references, which arises from differences in both linguistic content and emotional expression.
Directly applying prosodic features from a mismatched reference without accounting for these discrepancies can lead to unnatural or distorted speech.

In this work, we propose Maestro-EVC, a novel controllable EVC framework that harmonizes various attributes of emotional speech, including content, speaker identity, emotion, and temporal dynamics. 
We achieve controllability by effectively disentangling content, speaker, and emotion information from separate reference utterances, allowing each attribute to be independently controlled.
We also introduce a temporal emotion representation and explicitly model the prosody of the emotion reference even under prosody-mismatched conditions, thereby enabling the transfer of target temporal emotional dynamics.
Specifically, we first propose temporal content-aware emotion modeling ({\conemo}), which leverages a cross-attention mechanism~\cite{vaswani2017attention} to generate linguistic structure-aware temporal emotion embeddings. It allows the model to capture temporally fine-grained emotional dynamics from the emotion reference.
Second, we present explicit emotion prosody transfer ({\emopro}), incorporating a prosody augmentation strategy that simulates prosody-mismatched conditions during training, resulting in more robust prosody modeling.
Finally, we introduce the emotion-invariant speaker encoder ({\spkenc}), where emotional information in speaker embeddings is suppressed using a gradient reversal layer (GRL)~\cite{ganin2015unsupervised}, and speaker consistency is further reinforced via a triplet loss.
As a result, {\sysname} achieves high-quality emotional voice conversion that exhibits both controllability and accurate emotional expressiveness, guided by reference inputs.

Our contributions are summarized as follows:
\begin{itemize}
\item We propose {\sysname}, a controllable EVC framework that independently controls linguistic content, speaker identity, and emotional style using three distinct references.
\item We introduce a temporal emotion representation and an explicit prosody modeling method to capture and transfer temporally fine-grained emotional styles, even under prosody-mismatched conditions.
\item Through objective and subjective evaluations, we demonstrate that our method generates high-quality speech with rich emotional expressiveness and accurate control over each target attribute.
\end{itemize}

Audio samples are available at \url{https://maestroevc.github.io/demo/}.

%% file: Sections/2_Methods.tex
\section{Methods}

Fig. 1 illustrates the overall architecture of {\sysname}. 
During inference, the model takes three reference utterances for content, emotion, and speaker identity, which are encoded into latent representations.
A cross-attention mechanism combines the content and temporal emotion style representations to produce a content-aware emotion embedding. 
Target duration is predicted using this embedding and the duration representation from the emotion reference.
The FE (F0/Energy) predictor receives F0 and energy extracted from the emotion reference, along with content and temporal emotion representations.
The predicted content, emotion, prosody, and speaker representations are integrated and fed into a HiFi-GAN~\cite{kong2020hifi} vocoder for waveform synthesis.
In the following subsections, we provide a detailed description of each component of {\sysname}.

\subsection{Content Encoder}
\label{sec:content_encoder}
To extract a content representation that captures only linguistic information from input reference audio, we follow prior works~\cite{li2024sefvcspeakerembeddingfree, kreuk2022textlessspeechemotionconversion} that utilize a pre-trained HuBERT model~\cite{hsu2021hubertselfsupervisedspeechrepresentation}, which was trained with a masked prediction task on audio signals.
Given a content reference $x_c$, the HuBERT model encodes it into a sequence of frame-level continuous representations $z$.
To discretize $z$, we apply K-means clustering to obtain a sequence of discrete units $\hat{z}$, which are then mapped to learnable embeddings via an embedding table, resulting in the discrete content representation $c$.

\subsection{Temporal Content-aware Emotion Modeling (TCEM)}
\label{sec:temporal_content_aware_emotion_modeling}
To achieve temporally fine-grained emotional style transfer, we extract emotion representations at the frame level and align them with the target content via cross-attention mechanism.
Assuming the resulting representations may contain unintended content information, we apply a gradient reversal layer (GRL) to the cross-attention output to suppress residual content cues.


\subsubsection{Temporal Emotion Encoder}
To extract fine-grained temporal emotion representation, we adopt the approach of Wang \textit{et al.}~\cite{wang2023speechemotiondiarizationemotion}, which formulates speech emotion diarization as a task of predicting both emotion labels and their frame-level boundaries.
We use pre-trained model which has proven effective in downstream tasks such as emotional speech synthesis.

\subsubsection{Cross-attention Mechanism}


We use a cross-attention mechanism to temporally align frame-level emotional cues with the separately encoded linguistic content. 
Given the content and emotion references, $x_c$ and $x_e$, the content encoder produces a frame-level sequence $c$, while the temporal emotion encoder generates a sequence of emotion embeddings $e$.
To incorporate the emotional information into the linguistic content in a content-aware manner, we use $c$ as the query sequence $Q$, and $e$ as the key and value sequence $K$ and $V$, respectively, resulting in an aligned emotion representation $\hat{e}$.

\subsubsection{Residual Content Disentanglement}
Although the temporal emotion encoder is trained to extract frame-level emotion representations, its short-term acoustic inputs can inherently contain both emotional and phonetic information.
We hypothesize that this feature-level entanglement causes the resulting representation $\hat{e}$ to retain unintended linguistic cues.
This can yield prosodic artifacts, degrading both the naturalness and emotional expressiveness, especially when transferring emotion across mismatched linguistic content.

To mitigate this, we apply a projection block to the cross-attention output, followed by a GRL and content classifier during training.
The content classification loss $\mathcal{L}_{cont}^{GRL}$ is imposed adversarially through the GRL to suppress residual linguistic information in the emotion representation.
This yields a disentangled emotion representation $\hat{e}_d$ that effectively preserves fine-grained emotional style from $x_e$ while being temporally aligned with the target content.

Ablation results presented in Table~\ref{table_1} empirically support our hypothesis.
Removing the content GRL reduces both emotional expressiveness and content fidelity, indicating that residual linguistic cues in the emotion representation interfere with effective style transfer.

\subsection{Explicit Emotional Prosody Transfer (EEPT)}
To explicitly transfer the F0 and energy of the target prosody, we apply smoothing and prosody augmentation to these features and use them as conditions for the FE predictor, which generates the predictions aligned with the target content. 

\subsubsection{Prosody Extractor}
We first extract three prosodic features from the emotion reference: F0, energy, and duration denoted as $\nu$, $\eta$, and $d$, respectively.
To emphasize the overall contour of prosodic patterns while suppressing micro-level fluctuations, we apply a Savitzky-Golay filter~\cite{savitzky1964smoothing} to smooth the extracted features. This step ensures that prosody transfer relies on general prosodic trends rather than on content-specific perturbations. The smoothed F0, energy and duration are denoted as $\nu_s$, $\eta_s$, and $d_s$.

\subsubsection{Prosody Augmentation}
\label{sec:prosody_augmentation}
Reconstruction-based framework constrains the content and emotion reference to have the same prosody.
Thus, it cannot directly learn from prosody-mismatched scenarios, often resulting in unnatural speech during inference.
To address this, we introduce a prosody augmentation strategy that enables indirect learning of prosody transfer under prosody-mismatched conditions. 

During training, $\nu_s$ and $\eta_s$ are randomly augmented using either random shifting or piecewise time warping, each selected with equal probability.
Random shifting, which shifts the entire prosody sequence along the time axis by a random amount, simulates misalignment between content and prosody preserving the internal prosodic pattern. 
Piecewise time warping segments the prosody sequence, randomly stretches or compresses each segment along the time axis, and then concatenates and rescales the result to the original length.
This simulates partial mismatches in speaking rate or rhythm.
Formally, the augmentation process is defined as:
\begin{equation}
\nu_a, \eta_a = \mathrm{ProAug}\left(\nu_s, \eta_s\right),
\end{equation}
where $\nu_a$ and $\eta_a$ are the augmented F0 and energy, and $\mathrm{ProAug}(\cdot)$ denotes the prosody augmentation module. 
These augmentations allow the model to explicitly transfer prosody from any prosody-mismatched reference pair, preserving naturalness and expressiveness of speech during inference.

\subsubsection{FE Predictor}
\label{prosody_predictor}
To enable explicit prosody transfer adapted to the target content, we leverage not only the augmented F0 and energy but also incorporate the content embedding and the voiced/unvoiced (VUV) mask of the content reference $x_c$.

The VUV information plays a crucial role in guiding the model toward the perceptually relevant regions for prosody transfer. 
As F0 and energy have limited relevance in unvoiced segments, explicitly incorporating the VUV mask enables the model to assign the essential prosodic information from the emotion reference $x_e$ to the voiced regions of $x_c$.

The inputs to the FE predictor are formally defined as follows:
\begin{equation}
\hat{\nu}, \hat{\eta} = \mathrm{FEPred}\left(\nu_e + \eta_e + c + v \right),
\end{equation}

where $\hat{\nu}$ and $\hat{\eta}$ denote the predicted F0 and energy, $\nu_e$ and $\eta_e$ are the F0 and energy projected into a shared embedding space, $c$ and $v$ denote the discrete content representation and VUV mask extracted from $x_c$, and $\mathrm{FEPred}(\cdot)$ denotes the FE predictor.

\subsubsection{Duration Predictor}

To incorporate the target duration patterns, we predict the unit durations of $x_c$ based on its unique unit sequence, the smoothed durations $d_s$ and disentangled emotion representation $\hat{e}_d$ derived from $x_e$. 
We design the duration predictor to take these three inputs for estimating the duration of each unit.

We first extract sequences of discrete units, $\hat{z}_c$ and $\hat{z}_e$ from $x_c$ and $x_e$, respectively.
To obtain a distinct sequence of units and their corresponding repetition counts, we apply a deduplication operation:
\begin{equation}
\hat{z}_{uniq}, n_{count} = \text{dedup}(\hat{z}),
\end{equation}
where $\hat{z}_{uniq}$ denotes the sequence of unique units, and $n_{count}$ indicates the number of consecutive occurrences for each unit, which serves as the duration $d$.

From $x_e$, we extract $n_{count}$, which is then smoothed using a Savitzky–Golay filter to obtain the $d_s$.
We also extract $\hat{e}_d$ from $x_e$.
These, together with the unique unit sequence from $x_c$, are fed into the duration predictor to estimate the predicted duration $\hat{d}$.

\subsubsection{Prosody Loss}
During training, the FE predictor and duration predictor are optimized to predict their respective ground-truth targets. Specifically, the FE predictors are trained to estimate $\nu$ and $\eta$, while the duration predictor learns to predict $d$ extracted from $x_c$, as formulated below:

\begin{equation}
\mathcal{L}_{prosody} = \mathcal{L}_{f0} + \mathcal{L}_{energy} + \mathcal{L}_{dur},
\end{equation}
where $\mathcal{L}_{f0}$ and $\mathcal{L}_{energy}$ are L2 losses for F0 and energy prediction, and $\mathcal{L}_{dur}$ is the L1 loss for duration prediction.

\setlength{\textfloatsep}{0pt}
\setlength{\intextsep}{0pt}
\setlength{\tabcolsep}{4pt}
\begin{table*}[!t]
\centering
\normalsize
\caption{Objective evaluation results for WER, CER, EECS, SCA, F0-PCC, and E-PCC.}
\label{table_1}
\vspace{-2pt}
\begin{tabular}{lcccccc}
\toprule
\multicolumn{1}{l}{Model} & WER(\%)$\downarrow$ & CER(\%)$\downarrow$ & EECS$\uparrow$ & SCA(\%)$\uparrow$ & F0-PCC$\uparrow$ & E-PCC$\uparrow$ \\
\midrule
StyleVC~\cite{du2021disentanglement} & 16.79 & 9.46 & 0.537 & 90.10 & 0.380 & 0.297 \\
ZEST~\cite{dutta2024zero} & 17.18 & 9.85 & 0.779 & 93.54 & 0.432 & 0.293 \\
{\sysname} (Ours) & \textbf{11.78} & \textbf{6.54} & \textbf{0.819} & \textbf{93.69} & \textbf{0.551} & \textbf{0.316} \\
\midrule
\quad w/o content GRL              & 20.98 & 12.38 & 0.771 & 86.80 & 0.501 & 0.312 \\
\quad w/o temporal emotion representation   & 12.37 & 7.13 & 0.812 & 77.25 & 0.549 & 0.283 \\
\quad w/o Prosody Augmentation   & 17.56 & 10.37 & 0.786 & 88.81 & \textbf{0.566} & \textbf{0.336} \\
\quad w/o $\mathcal{L}_{spk}$             & 12.56 & 7.11 & 0.773 & 89.81 & 0.536 & 0.301 \\
\bottomrule
\end{tabular}
\vspace{-10pt}
\end{table*}

\subsection{Emotion-Invariant Speaker Encoder (EISE)}
We derive embeddings from the speaker reference $x_s$ that are invariant to emotional attributes while preserving speaker identity.
We adopt a pre-trained ECAPA-TDNN~\cite{Desplanques_2020}, widely used for robust speaker representations, though it may still encode emotional information.


In order to mitigate the entanglement between speaker identity and emotional information, we append trainable layers to the output of the frozen pre-trained ECAPA-TDNN, referred to as the speaker encoder.
A GRL and an emotion classifier are applied to the appended layers.
The speaker encoder is trained adversarially using the emotion classification loss $\mathcal{L}_{emo}^{GRL}$ reversed by the GRL to encourage the suppression of emotional cues in the appended layers.

Although the GRL discourages the encoder from retaining emotional information, it does not explicitly enforce consistency across embeddings of the same speaker under different emotional conditions.

To address this limitation, we incorporate a triplet loss based on cosine similarity, which encourages embeddings of the same speaker under different emotional states to be more similar than those of different speakers.
Each triplet consists of an anchor, a positive sample from the same speaker with different emotions, and a negative sample from a different speaker.
The loss is defined as:
\begin{align}
\mathcal{L}_{trip} = \sum_{i=1}^{N} \big[ 
& \operatorname{sim}\left(E_s(x_i^a), E_s(x_i^n)\right) \nonumber \\
& - \operatorname{sim}\left(E_s(x_i^a), E_s(x_i^p)\right) + \alpha 
\big]_+,
\end{align}
where $\operatorname{sim}(\cdot,\cdot)$ denotes the cosine similarity, and $x_i^a, x_i^p, x_i^n$ represent the anchor, positive, and negative samples, respectively.
The margin $\alpha$, set to 0.3, defines the minimum desired separation between the positive and negative pairs.
The total speaker loss is defined as:
\begin{align}
    \mathcal{L}_{spk} = \mathcal{L}_{trip} + \mathcal{L}_{emo}^{GRL},
\end{align}
By combining GRL and triplet loss, the speaker encoder is encouraged to suppress emotional information and to maintain speaker-consistent embeddings across emotions.

\subsection{Training strategy}\label{SCM}
The model is trained to reconstruct the input waveform. A single input $x$ serves as $x_c$, $x_e$, and $x_s$ with HiFi-GAN~\cite{kong2020hifi} as the vocoder.
The generator $G$ and discriminator $D$ are optimized with the following losses:
\begin{equation}
    \mathcal{L}_{G} = \mathcal{L}_{adv}(G; D) + \mathcal{L}_{fm} + \mathcal{L}_{recon}(G), 
\end{equation}
\begin{equation}
    \mathcal{L}_{D} = \mathcal{L}_{adv}(D; G),
\end{equation}
where $\mathcal{L}_{adv}$, $\mathcal{L}_{fm}$, and $\mathcal{L}_{recon}(G)$ represent the adversarial, feature matching, and reconstruction losses, respectively.
The total loss for $G$ is given by:
\begin{equation}
\mathcal{L}_{G}^{total} = \mathcal{L}_{G} + \mathcal{L}_{spk} + \mathcal{L}_{cont}^{GRL} + \mathcal{L}_{prosody},
\end{equation}
where $\mathcal{L}_{spk}$, $\mathcal{L}_{cont}^{GRL}$, $\mathcal{L}_{prosody}$ are auxiliary losses for speaker supervision, content disentanglement via GRL, and prosody modeling, each weighted by a tunable coefficient $\lambda$.

%% file: Sections/3_Experiments.tex
\section{Experiments}

\subsection{Experimental Setup}
\subsubsection{Dataset}
We used the 12-layer base HuBERT model~\cite{hsu2021hubertselfsupervisedspeechrepresentation} pre-trained on 960 hours of the LibriSpeech dataset~\cite{7178964}, and the ECAPA-TDNN pre-trained on the VoxCeleb dataset~\cite{nagrani2017voxceleb}.
For training and evaluation, we used the English partition of the Emotional Speech Dataset~\cite{zhou2022emotional}, which contains 350 parallel utterances at 16 kHz from 10 English speakers across five emotions: neutral, happy, angry, sad, and surprise.

\subsubsection{Implementation details} 
In our implementation, the content encoder used a vocabulary size of 500, with each token embedded into a 256-dimensional vector.
Input audio was converted to an 80-bin Mel-spectrogram with a window size 1,024 and a hop size 256, which was used for both Mel-based reconstruction loss and frame-level energy extraction.
Additionally, F0 was extracted using the WORLD vocoder~\cite{morise2016world}. 
In prosody augmentation, shifting moves the sequence by a random value in [-15, 15] frames, and piecewise time warping randomly splits it into 2–5 segments, each scaled by a factor randomly sampled from [0.4, 1.6].
Both the FE and duration predictors consist of two stacked Transformer blocks with 1D convolution layers replacing the feed-forward network~\cite{vaswani2017attention}.
The weight $\lambda_{recon}$ for Mel-based reconstruction loss was set to 45, while all other loss weights were set to 1.
The AdamW optimizer was used, with a learning rate of $2 \times 10^{-4}$.

\subsubsection{Baselines}
We adopt StyleVC~\cite{du2021disentanglement} and ZEST~\cite{dutta2024zero} as our baseline models. 
StyleVC is an any-to-any expressive voice conversion framework designed to disentangle linguistic content, speaker identity, pitch, and emotional style information, enabling simultaneous conversion of arbitrary speaker identity and emotional style. 
ZEST is a zero-shot EVC framework that separates speaker and emotion representations and predicts F0 from the extracted content, speaker, and emotion features, allowing it to handle reference-guided conversion with prosody transfer.
To the best of the authors’ knowledge, there has been no prior EVC model that simultaneously considers both pitch and energy in prosody modeling.
Therefore, we selected these two models as baselines for their focus on pitch transfer.

\subsection{Evaluation Settings}

Rather than restricting speaker reference to the neutral emotional state, this experiment employed emotionally expressive references to evaluate the conversion to the target emotion style, thus enabling a more comprehensive assessment.

\subsubsection{Seen dataset evaluation}
We randomly constructed 700 test input sets, each composed of a content, speaker, and emotion reference drawn from different speakers and containing distinct linguistic content.
This setting ensures diverse evaluation conditions.

\subsubsection{Zero-shot evaluation}
\label{unseen_scenario}
To assess generalization, we evaluated {\sysname} under two unseen scenarios: unseen speakers (US) using 18 speakers from the VCTK corpus~\cite{veaux2017vctk}, and unseen emotion states (UE) using held-out classes, fear and disgust from the CREMA-D~\cite{cao2014crema} and frustration and excitement from IEMOCAP~\cite{busso2008iemocap}.


\subsubsection{Evaluation metrics}
We evaluated the converted speech using six objective metrics.
For intelligibility, we computed word error rate (WER) and character error rate (CER) using Whisper~\cite{radford2023robust}.
Emotion similarity was measured by emotion embedding cosine similarity (EECS) with the emotion2vec+ base9 model~\cite{ma2023emotion2vec}.
Speaker similarity was assessed via speaker classification accuracy (SCA) based on pre-trained classifier.
We evaluated prosody alignment via Pearson correlation coefficient (PCC)~\cite{cohen2009pearson} between the F0 and energy trajectories of the synthesized speech and the reference, after aligning them using dynamic time warping~\cite{muller2007dynamic}.

For subjective evaluation, we conducted a Mean Opinion Score (MOS) test with 25 human participants, using 30 randomly sampled test pairs.
Participants rated naturalness, emotional similarity, speaker similarity, and prosody similarity, by comparing each sample with the corresponding target reference.
For prosody, they assessed the similarity of temporal variations in pitch, intensity, and speaking rate.


%% file: Sections/4_Results.tex
\section{Results}
\subsection{Objective Evaluations}
As shown in Table~\ref{table_1}, under the seen scenario, {\sysname} outperforms both baselines across all objective metrics.
Higher PCCs for F0 and energy indicate superior prosody modeling and faithful transfer of emotion reference.
These results confirm that {\sysname} achieves controllability through effective disentanglement of each attribute, while accurately modeling the temporal dynamics of emotional expression.

Table~\ref{table_2} shows that, in zero-shot tests on unseen speakers and unseen emotions, {\sysname} consistently surpasses the baselines on all metrics.
These results confirm the strong generalization capability of {\sysname} to both unseen speakers and emotion states.


\setlength{\tabcolsep}{2pt}
\setlength{\textfloatsep}{0pt}
\setlength{\intextsep}{0pt}
\setlength{\floatsep}{0pt}
\begin{table}[!t]
\small
\centering
\caption{Objective evaluation results in unseen scenarios}
\label{table_2}
\vspace{-2pt}
\begin{tabular}{llcccc}
\toprule
Scenario & Model & CER(\%)$\downarrow$ & EECS$\uparrow$ & SCA(\%)$\uparrow$ & F0-PCC$\uparrow$ \\
\midrule
\multirow{3}{*}{UE} 
& StyleVC~\cite{du2021disentanglement} & 10.13 & 0.575 & 85.70 & 0.310 \\
& ZEST~\cite{dutta2024zero}           & 11.01 & 0.692 & 88.00 & 0.313 \\
& {\sysname}                         & \textbf{9.64} & \textbf{0.768} & \textbf{88.67} & \textbf{0.370} \\
\midrule
\multirow{3}{*}{US} 
& StyleVC~\cite{du2021disentanglement} & 8.85 & 0.524 & - & 0.388 \\
& ZEST~\cite{dutta2024zero}           & 9.69 & 0.802 & - & 0.486 \\
& {\sysname}                          & \textbf{6.10} & \textbf{0.841} & - & \textbf{0.577} \\
\bottomrule
\end{tabular}
\end{table}

\setlength{\textfloatsep}{5pt}
\setlength{\intextsep}{0pt}
\begin{table}[!t]
\small
\centering
\caption{Subjective evaluation results in terms of MOS}
\label{table_3}
\vspace{-2pt}
\begin{tabular}{lcccc}
\toprule
Model & Naturalness$\uparrow$ & Emo.Sim.$\uparrow$ & Spk.Sim.$\uparrow$ & Pro.Sim.$\uparrow$ \\
\midrule
StyleVC~\cite{du2021disentanglement} & 3.88 $\pm$ 0.16 & 2.37 $\pm$ 0.17 & 3.91 $\pm$ 0.13 & 2.03 $\pm$ 0.15 \\
ZEST~\cite{dutta2024zero} & 3.54 $\pm$ 0.13 & 3.81 $\pm$ 0.14 & 3.46 $\pm$ 0.19 & 2.86 $\pm$ 0.17 \\
{\sysname} & \textbf{4.06 $\pm$ 0.12} & \textbf{4.11 $\pm$ 0.09} & \textbf{4.02 $\pm$ 0.11} & \textbf{4.15 $\pm$ 0.06} \\
\bottomrule
\end{tabular}
\end{table}

\subsection{Subjective Evaluations}
The results of subjective evaluations are presented in Table~\ref{table_3}.
Across all criteria, {\sysname} attains the highest scores, significantly surpassing both baselines.
In particular, {\sysname} shows a clear advantage in prosody similarity, indicating that explicit prosody modeling contributes significantly to the perception of expressiveness. 
These results indicate that {\sysname} not only improves objective quality but also delivers superior perceptual performance in EVC.

\subsection{Ablation Study}

To investigate the effect of our proposed methods in {\sysname}, we conduct an ablation study on three key components: (1) content classifier and GRL in the TCEM module, (2) temporal emotion representation, (3) prosody augmentation in the EEPT module, and (4) the speaker loss $\mathcal{L}_{spk}$ in EISE.
The results are summarized in Table~\ref{table_1}.

First, we investigate the effect of content disentanglement in the TCEM module by removing the content classifier and GRL. 
This ablation causes significant degradation across all metrics, especially in WER and CER, indicating that content leakage in the emotion representation impairs reconstruction under linguistic mismatch.
Adversarial training therefore improves emotional expressiveness and content preservation.

Second, we evaluate the effect of temporal emotion representation by replacing the temporal emotion encoder with pre-trained utterance-level emotion encoder.
The results show decreases across all metrics, with particularly large drops in SCA and E-PCC.
These findings indicate that the temporal emotion representation contributes to notable improvements in various aspects of performance.

Third, to assess the impact of prosody augmentation, we remove the augmentation step during training. 
This yields notable drops in WER, CER, EECS, and SCA, indicating reduced robustness under prosody mismatches.
Without exposure to prosodic variation, the model becomes overly dependent on the reference and attempts to forcibly align mismatched prosodic patterns with the source content. 
This often leads to unclear articulation and unnatural temporal dynamics, where the rhythm and emphasis fail to align with the linguistic structure.
Consequently, although F0-PCC and E-PCC slightly increase as the model rigidly follows the reference, overall naturalness and generalization decline.
These results validate prosody augmentation for robust, natural prosody transfer.

Lastly, we evaluate the speaker encoder by removing the speaker loss $\mathcal{L}_{spk}$.
This ablation lowers SCA and slightly reduces EECS, suggesting residual emotional information in the speaker embedding that hinders accurate target-emotion modeling.
It highlights the need to minimize emotional entanglement and maintain embedding consistency for reliable identity control.


\begin{figure}[t]
  \centering
  \includegraphics[width=0.48\linewidth]{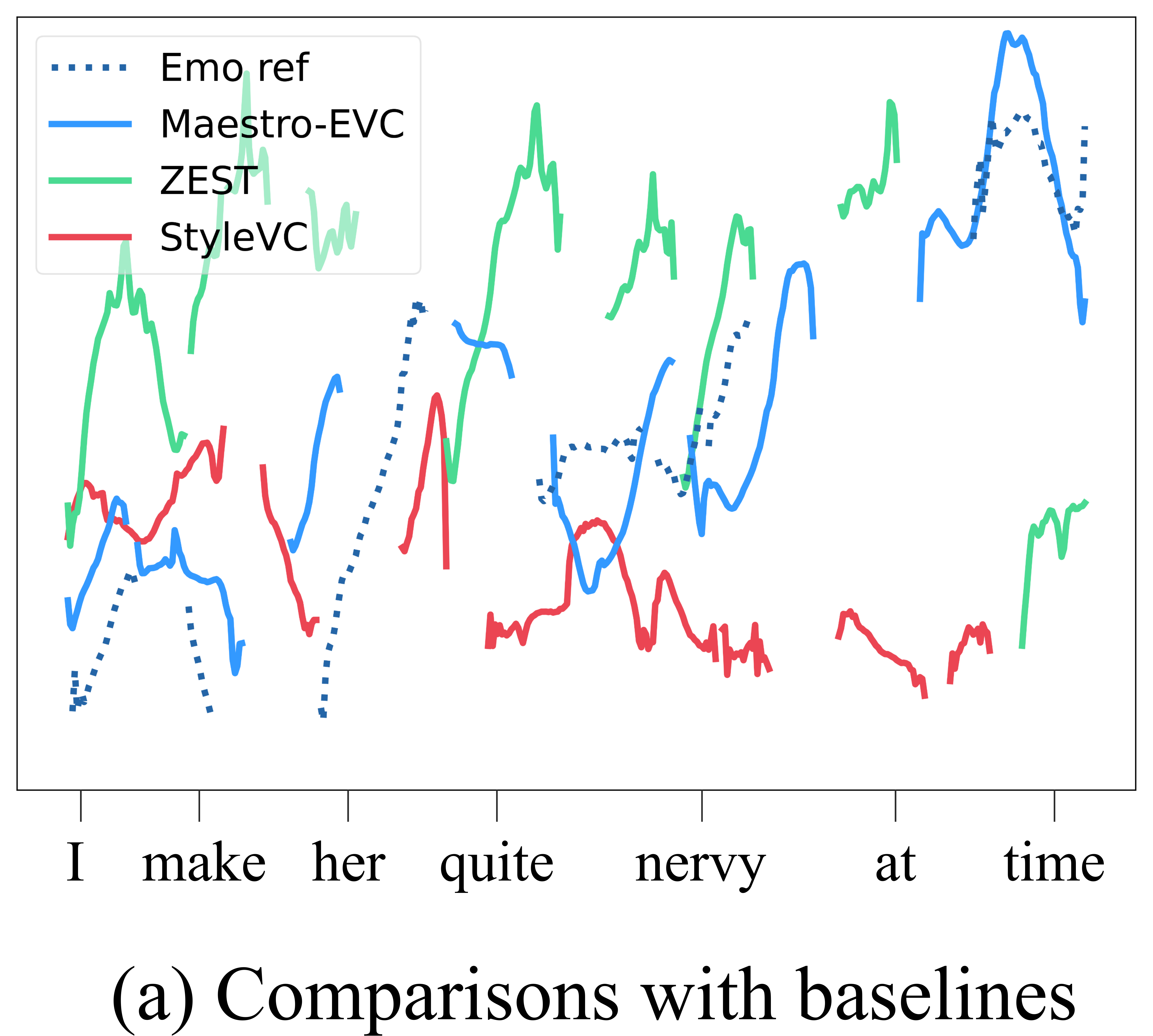}
  \hspace{-0.01\linewidth}
  \includegraphics[width=0.48\linewidth]{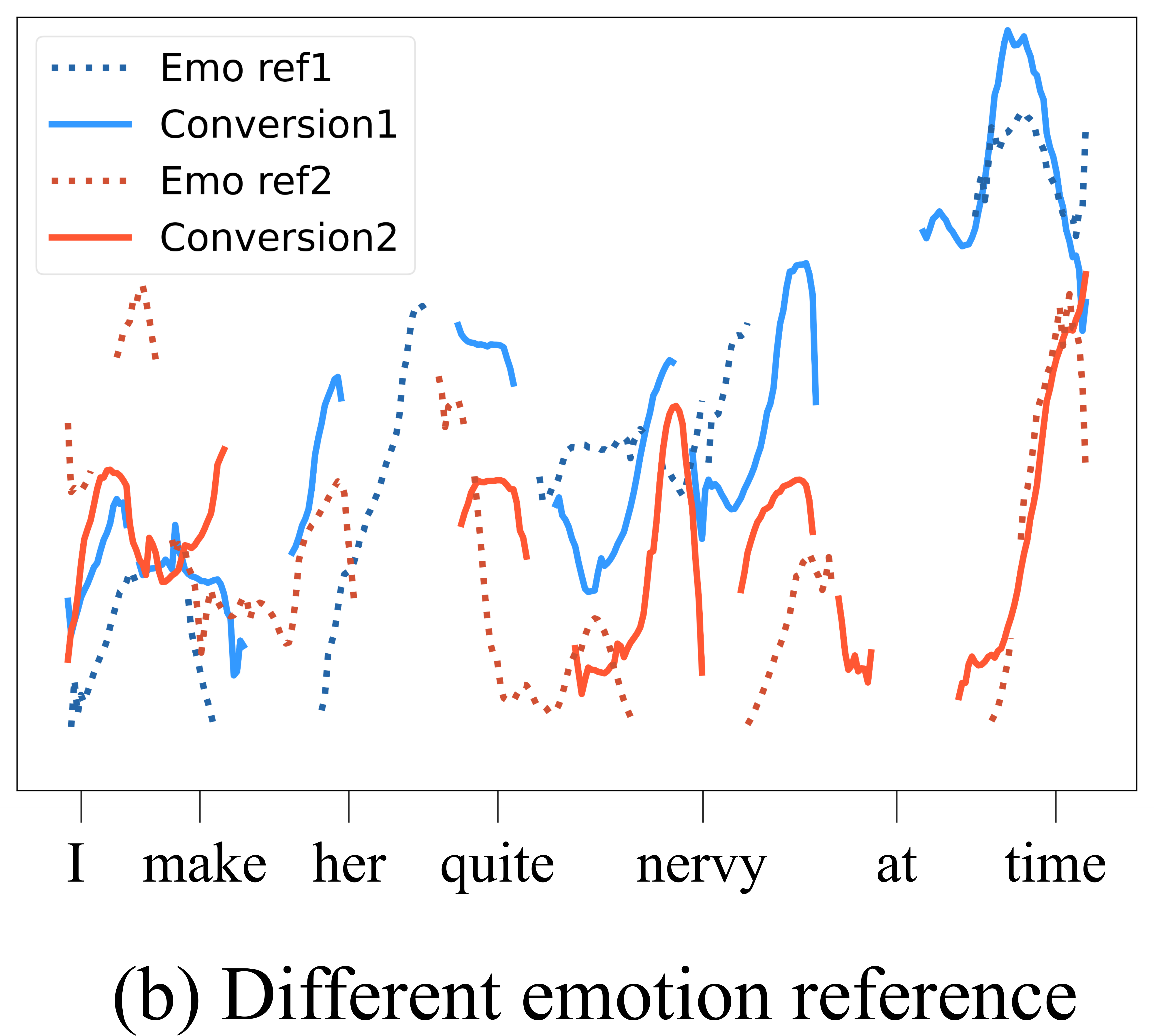}
  \vspace{-0.5em}
  \caption{Visualization of F0 contours. (a) shows F0 comparisons with baselines, while (b) shows the results of Maestro-EVC using different emotion references. In all conversions, the content and emotion references differ in both emotion category and linguistic content. The two curves in (b) correspond to conversions using different utterances from the ``Surprise'' category as emotion references.}
  \label{fig:f0_comparison}
\end{figure}

\subsection{Explicit Prosody Transfer}
Fig.~\ref{fig:f0_comparison} shows the F0 contours of the emotion reference and the converted speech. 
As shown in (a), compared with the baseline models, {\sysname} more accurately follows the pitch contour of the emotion reference.
Furthermore, (b) shows that, even within the same emotion category, variations in prosodic expression across different references result in distinct outputs, indicating that our model effectively reflects fine-grained prosodic differences.


%% file: Sections/5_Conclusion.tex
\section{Conclusion}
In this paper, we propose {\sysname}, a novel controllable EVC framework that harmonizes various attributes of emotional speech, including content, speaker identity, emotion, and temporal dynamics.
By disentangling content, speaker, and emotion representations, it enables independent control of each attribute using separate reference utterances, even with any reference combination.
To achieve rich expressiveness, we introduce a temporal emotion representation and explicit prosody transfer, enabling effective performance even in prosody-mismatched scenarios.
Experimental results demonstrate that {\sysname} outperforms existing baselines across all metrics in both seen and zero-shot scenarios, validating its controllability and expressiveness.

%% file: Sections/6_Acknowledgment.tex
\section{Acknowledgment}
This research was partly supported by Basic Science Research Program through the National Research Foundation of Korea(NRF) funded by the Ministry of Education(RS-2022-NR070870), Institute of Information \& communications Technology Planning \& Evaluation (IITP) grant funded by the Korea government(MSIT) (No.RS-2019-II191906, Artificial Intelligence Graduate School Program(POSTECH)).